# Single step synthesis of $Sr_4V_2O_6Fe_2As_2$: The blocking layer based potential future superconductor

Anand Pal, Arpita Vajpayee, R. S. Meena, H. Kishan and V. P. S. Awana[*]

National Physical Laboratory, Dr. K.S. Krishnan Road, New Delhi-110012, India

We report synthesis, structural details and transport measurements on $Sr_4V_2O_6Fe_2As_2$. Namely the stoichiometric amounts of $V_2O_5$ + 1/2×$SrO_2$ + 7/2×Sr + 2×FeAs are weighed mixed, ground thoroughly and palletized in rectangular form in a glove box in high purity Ar atmosphere. The pellet is further sealed in an evacuated ($10^{-5}$ Torr) quartz tube and put for heat treatments at 750 and 1150 $^0$C in a single step for 12 and 36 hours respectively. Finally the quartz ampoule is allowed to cool naturally to room temperature. The as synthesized sample is black in color. The compound crystallized in *P*4/*nmm* space group with lattice parameters a = b = 3.925 Å and c = 15.870 Å. Also seen are some small impurity lines. The compound did not exhibit superconductivity but instead a spin density wave (*SDW*) like metallic step at around 175 K is seen in *R*(*T*) measurements. Principally in $[FeAs]^{-1}\{Sr_4V_2O_6\}^C[FeAs]^{-1}$ the net value of blocking layer charge C must be either less or more than 2, to let it be electron or hole type superconductor respectively. Efforts are underway to achieve superconductivity in the studied system.

[*]*Corresponding Author:*

*Dr. V.P.S. Awana*

*Fax No. 0091-11-25626938: Phone no. 0091-11-25748709*

*e-mail-awana@mail.nplindia.ernet.in: www.freewebs.com/vpsawana/*

The recent discovery of superconductivity in oxy-pnictides viz REFeAsO (1111) and $MFe_2As_2$ (122) with their superconducting transition temperature ($T_c$) of up to 60 K has attracted a lot of attention [1-5]. Primarily it seems that the evolution of superconductivity in the so-called 1111 and 122 oxy-pnictides originates from a magnetic state, much alike the cuprate superconductors [3-8]. Although the anti-ferromagnetic ground state of these new oxy-pnictides is yet not established unambiguously, the short-range magnetic order in terms of spin density wave (*SDW*) are seen before onset of superconductivity [1-5]. It is difficult to achieve stoichiometric ground state of either 1111 phase or the 122 primarily due to non-stabilization of REFeAsO structure with full oxygen stoichiometric in the previous case and without the presence of defect structure in second. For example, it is found repeatedly that oxygen stoichiometric REFeAsO being synthesized by closed cell *HPHT* (high pressure high temperature) is never a single phase compound, but always accompanied with some impurity [9]. Hence till very recently, the ground state compounds of these oxy-pnictides were really warranted. In this direction some research has taken place in last couple of weeks and a new series of compounds viz. $A_4M_2O_6Fe_2As_2$ with A = Ca, Sr Ba, and M = V, Cr, or Sc has been reported [10-14]. These compounds are nicknamed as 42622. Principally the perovskite $Sr_4M_2O_6$ block is inserted between FeAs layers with charge neutrality as $[FeAs]^{-1}\{Sr_4M_2O_6\}^{+2}[FeAs]^{-1}$. Or otherwise the perovskite block is inserted between two superconducting FeAs layers. Now for the ground state of the compound the perovskite block with $\{Sr_4M_2O_6\}^{+2}$ does require the M cation to be exactly in trivalent state. Interestingly this way one can, realize the ground state of these compounds. Further one finds a lot of room for transfer of mobile carriers from perovskite block to adjacent possible superconducting FeAs layers by changing the charge neutrality of the blocking/redox block.

In the short letter we report the successful synthesis of the important very recently invented $Sr_4V_2O_6Fe_2As_2$ (42622) compound by an easy and versatile single step synthesis method via vacuum encapsulation technique. The compound is crystallized in near single phase in *P*4/*nmm* space group but with few impurities. As expected the compound did not exhibit superconductivity but instead a spin density wave (*SDW*) like metallic step at around 175 K. This is evidenced in *R*(*T*) measurements. We believe the 42622 phase has lot of scope for control of superconductivity and higher $T_c$ values. The possibilities are discussed in the short letter.

First the Fe (Alfe Aesar, 99.2%) and As (Alfe Aesar, 99%) chips were sealed in evacuated quartz tube and heat-treated for 12 hours at 700$^o$C. Later, the stoichiometric amounts of $V_2O_5$ (Aldrich, 99.6%) + 1/2×$SrO_2$ (Aldrich) + 7/2×Sr (Alfa Aesar, 99%) + 2×FeAs are weighed, mixed, ground thoroughly and palletized in rectangular form in a glove box in high purity Ar atmosphere. The pellet is further wrapped in tantalum foil and then sealed in an evacuated ($10^{-5}$ Torr) quartz tube and put for heat treatments at 750 and 1150 $^0$C in a single step for 12 and 36 hours respectively. Finally the quartz ampoule is allowed to cool naturally to room temperature. The photograph of sample containing sealed quartz capsule is shown in inset of Fig. 1. The as synthesized sample is black in color. The x-ray diffraction patterns of these compounds were taken on Rigaku mini-flex II diffrractometer and intensity data were collected in the 2θ range of 20° - 70° at a step of 0.02° using Cu-*K*α radiation. The resistivity measurements were carried out by four-probe method on a close cycle refrigerator in temperature range of 20 to 300 K.

The room temperature X-ray diffraction (*XRD*) pattern for the $Sr_4V_2O_6Fe_2As_2$ sample and its Rietveld analysis are shown in Figure l. Inspection of the diffraction profile at room temperature reveals that the compound possesses the tetragonal unit cell with space group *P*4/*nmm*. It can also be seen from the figure that besides the majority phase, few impurity lines having low intensity are also seen in the *XRD* pattern. These extra lines are due to $Sr_2VO_4$ and FeAs phases. Rietveld analysis of the room temperature diffraction pattern proceeded smoothly and the lattice parameters are found to be: $a$ = 3.925(4) Å and c = 15.870(2) Å. These values of lattice parameters are in confirmation with earlier reports [10, 13]. Typical Rietveld refinement parameters for $Sr_4V_2O_6Fe_2As_2$ sample are given in Table 1.

The Resistance versus temperature (*R*-*T*) plot for the $Sr_4V_2O_6Fe_2As_2$ sample is shown in Figure 2. The *R*-T behavior is comparatively less metallic with a relatively high resistance around 142 mΩ at room temperature, which decreases slightly as the temperature goes down to around 175 K. Below 175 K the $Sr_4V_2O_6Fe_2As_2$ exhibits a metallic step down to the temperature of 55 K, i.e. resistance decreases monotonically with temperature and without any sign of superconductivity. Below 55 K, the resistance increases with temperature resulting to a semiconductor like behavior down to lowest studied temperature of 20 K. The compound did not exhibit superconductivity but instead a spin density wave (*SDW*) like metallic step at around 175 K is seen in *R*(*T*) measurements. The *R*(*T*) step metallic step in electrical conduction of oxy-pnictides is known to be as an indication of the spin density wave (*SDW*) transition of these systems [1-8,15]. The Absence of superconductivity in this compound is considered to be due to insufficient carrier concentration the same as in the case of undoped REFeAsO [1-5,15].

However there is an interesting difference, in REFeAsO the carriers are provided from non stoichiometric $[REO]^{+1}$ to lying below $[FeAs]^{-1}$ by charge non stoichiometry in the former via RE, O site substitution or oxygen deficiency. On the other hand in the presently studied $A_4M_2O_6Fe_2As_2$ system the charge reservoir works as a blocking layer between two superconducting FeAs layers [10-14]. The situation is similar to the HTSc M-1222 ($MSr_2Y_{2-x}Ce_xCu_2O_{10}$: x = 0.0 to 1.0 and M = Nb, Ru, Cu, Pb etc.) system [16], where a three-layer fluorite RE-Ce-$O_2$ block is being sandwiched between two superconducting Cu-$O_2$ layers [16]. Interestingly, the $A_4M_2O_6Fe_2As_2$ has many more possibilities to manipulate the superconductivity in adjacent FeAs layer. In stoichiometric compound the transition metal M is trivalent and A is divalent alkali metal, in this situation the blocking layer $A_4M_2O_6$ accounts for net charge = (8+6)-12, i.e. +2. This is exactly compensated with –2 charge from two $[FeAs]^{-1}$ layers. However a slight change in the stoichiometry such as $A^{+2}$ site $Na^{+1}$ or $K^{+1}$, O site $F^{-1}$or $O_{1-\delta}$ and a huge scope on M site with various other 3d transition metals substitution could possibly lead to various new superconducting compounds. In this context we believe the perovskite block inserted oxy-pnictides have huge future scope for invention of new superconductor in this class of materials. Although $Sr_4V_2O_6Fe_2As_2$ (42622) is reported [14], superconducting at above 37 K, we believe that is due to possible compositional non-stoichiometry, because otherwise as shown above the un-doped compound be an insulator. Even presently studied $Sr_4V_2O_6Fe_2As_2$ (42622) is also not an anti-ferromagnetic (*AFM*) insulator but an *SDW* material with moderately high resistivity. This we again conclude to be due to non-stoichiometric composition. In both ref. 14 and present study the impurity lines are seen in *XRD* and thus imposing compositional variation to the pristine $Sr_4V_2O_6Fe_2As_2$ (42622). It seems not only more work is needed to establish bulk superconductivity in the system by various onsite substitutions but the pristine *AFM* is also yet warranted similar to that as for 1111 and 122 phases. None the less the new $Sr_4V_2O_6Fe_2As_2$ (42622) phase has lot more scope to advance the research on oxy-pnictides in near future.

Very recently invented 14] new layered iron oxypnictides $Sr_4V_2O_6Fe_2As_2$ (42622) has been synthesized by an easy solid-state reaction method in vacuum with out employing the *HPHT* process. In resistivity measurement, the compound did not exhibit superconducting transitions probably due to insufficient carrier concentration, but a *SDW* state below 175K. Further attempts to introduce effective carriers to these materials must be promising for achieving high-*T*c superconductivity.

The authors thank Prof. Vikram Kumar, Director of the National Physical Laboratory, New Delhi, for his constant support and encouragement. Dr. M. Deepa from NPL is acknowledged use of the glove box. Anand Pal and Arpita Vajpayee would like to acknowledge *CSIR*, India for providing research fellowships. V P S Awana is also thankful to various fruitful discussions in regards to search for new oxy-pnictide superconductors with Prof. G. Baskaran from Mat. Sci. Inst. Chennai, India.

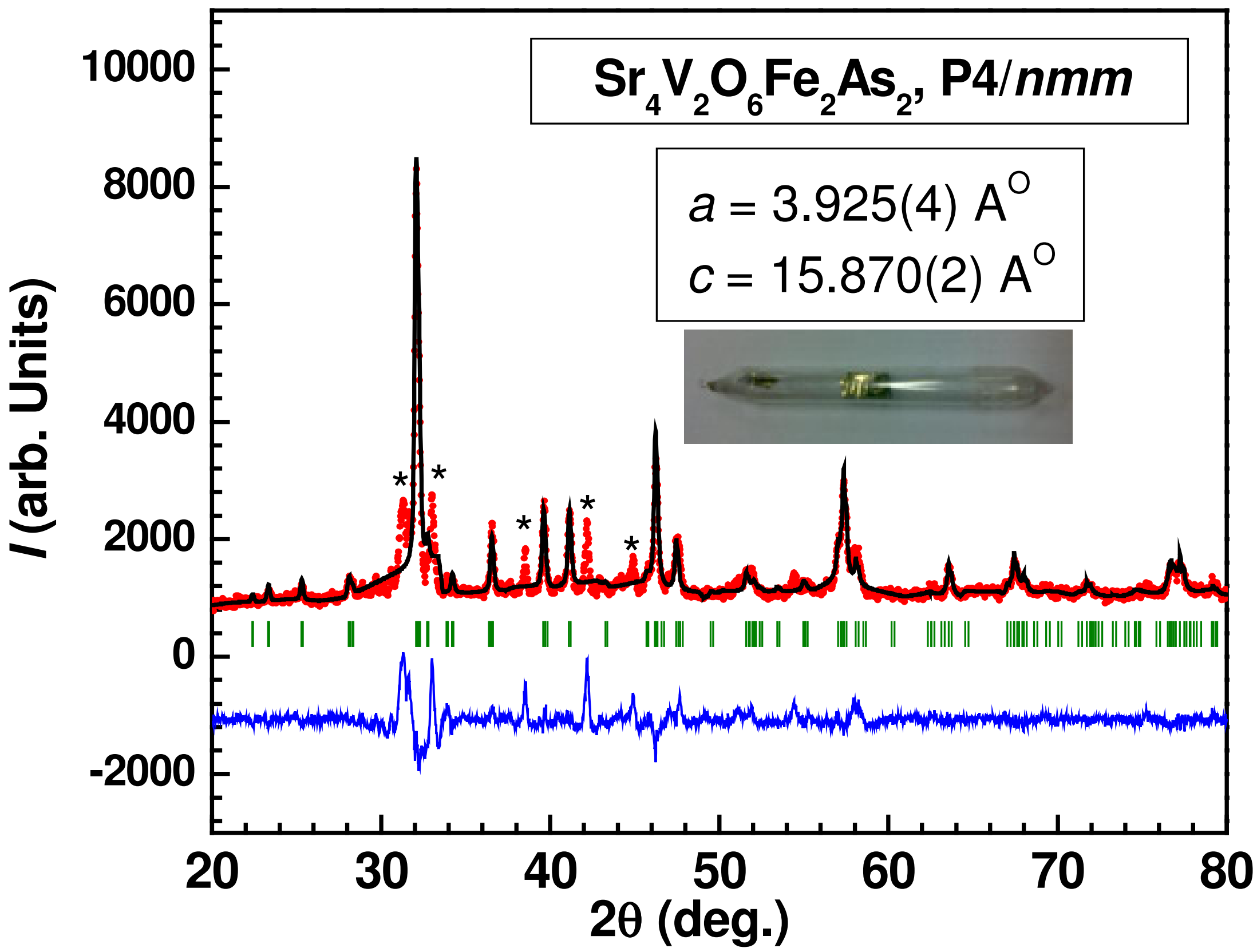

Figure 1. Room temperature fitted and observed *XRD* pattern of $Sr_4V_2O_6Fe_2As_2$ compound, The diffraction lines not permitted in the structural space group are marked with (*). These are secondary phase impurities in the matrix. The inset shows the sealed ampoule containing the sample wrapped in tantalum foil

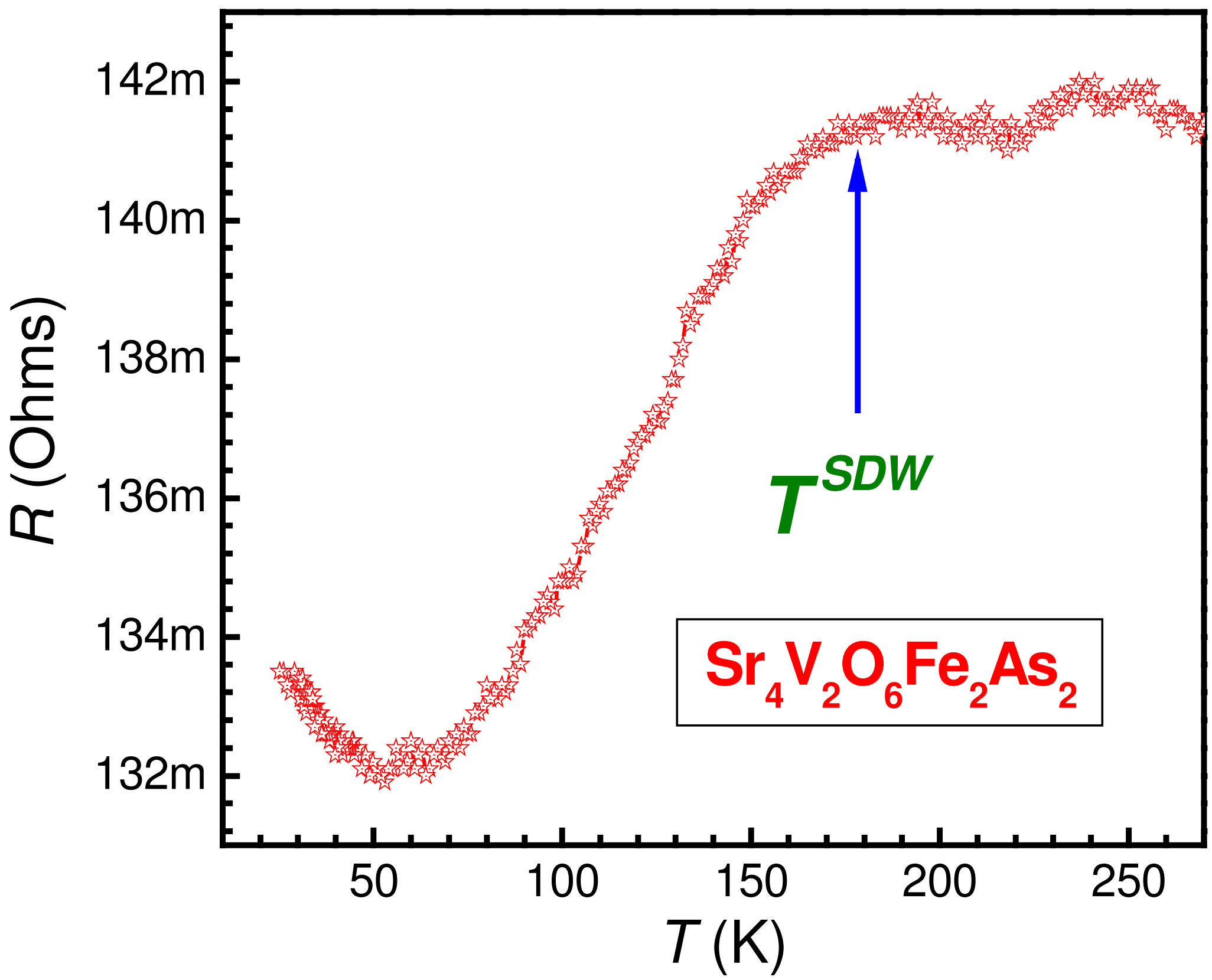

Fig. 2 Resistance versus Temperature *R*(*T*) of $Sr_4V_2O_6Fe_2As_2$ compound. The metallic step temperature is marked as the onset of spin density wave (*SDW*) transition.

Table 1. Rietveld refined parameters for $Sr_4V_2O_6Fe_2As_2$ compound, Space group: *P4/nmm*

| Atom | x | y | z |
|---|---|---|---|
| Sr1 | -0.250 | -0.250 | 0.1944( 14) |
| Sr2 | -0.250 | -0.250 | 0.4084( 14) |
| V | 0.25 | 0.25 | 0.3121( 20) |
| Fe | 0.250 | -0.250 | 0.0000 |
| As | 0.250 | 0.250 | 0.0909( 17) |
| O1 | 0.250 | -0.250 | 0.3121( 33) |
| O2 | 0.250 | 0.250 | 0.4209( 44) |

$a$ = 3.925(4) Å and $c$ = 15.870(2) Å

*Rp*: 6.97%, *Rwp*: 10.30%, *Rexp*: 2.80%